\documentclass[conference]{IEEEtran} % US 
\usepackage{amsthm,amsmath,amssymb,color,graphicx,caption,subcaption,comment,tikz,url,latexsym} %amsthm
\usepackage{pgfplots}
\usepackage{verbatim}
\usetikzlibrary{decorations.markings}
\usepackage{cite}
\usepackage{enumerate}
\usepackage[draft,pdfborder={0 0 0}]{hyperref}
\usepackage{colortbl}
\usepgfplotslibrary{units}
\usetikzlibrary{spy,backgrounds}
\usetikzlibrary{shapes.geometric} %for diamond in text
\usepackage{pgfplotstable}
\usepackage{flushend}
\usepackage[normalem]{ulem}
\usepackage{multirow}
\usepackage{nicefrac} %nicefrac
\usepackage{glossaries}
\newcommand{\stkout}[1]{\ifmmode\text{\sout{\ensuremath{#1}}}\else\sout{#1}\fi}
\theoremstyle{definition}

\newcommand{\PM}{\text{PM}}
\newcommand{\FM}{\text{FM}}

\newcommand{\Ical}{\mathcal{I}}
\newcommand{\Icalsort}{\mathcal{I}_{\text{sort}}}
\newcommand{\Lij}[2]{\lambda_{#1,#2}}

\newcommand{\Fcal}{\mathcal{F}}

\newcommand{\latSC}{\mathcal{L}_{\text{SC}}}
\newcommand{\latSCF}{\mathcal{L}_{\text{SCF}}}
\newcommand{\avgASCL}{\overline{\mathcal{L}_{\text{ASCL}}}}
\newcommand{\avgSCF}{\overline{\mathcal{L}_{\text{SCF}}}}
\newcommand{\avgSCLF}{\overline{\mathcal{L}_{\text{SCLF}}}}

\newcommand{\latSCL}{\mathcal{L}_{\text{SCL}}}

\newcommand{\avgPSCLF}{\overline{\mathcal{L}_{\text{PSCLF}}}}

\newcommand{\avgt}{\overline{t}}

\newcommand{\fixme}[2]{\ifx&#2&{\leavevmode\color{red}#1}\else{\leavevmode\color{red}FIXME\{}#1{\leavevmode\color{red}\}}\footnote{{\leavevmode\color{red}#2}}\PackageWarning{Fixme}{#1: #2}\fi}

\definecolor{matlab1}{rgb}{0.000, 0.447, 0.741} % blue
\definecolor{matlab2}{rgb}{0.850, 0.325, 0.098} % red
\definecolor{matlab3}{rgb}{0.929, 0.694, 0.125} % yellow
\definecolor{matlab4}{rgb}{0.494, 0.184, 0.556} % violet
\definecolor{matlab5}{rgb}{0.466, 0.674, 0.188} % green
\definecolor{matlab6}{rgb}{0.301, 0.745, 0.933} % light blue
\definecolor{matlab7}{rgb}{0.635, 0.078, 0.184} % dark purple

\newacronym{awgn}{AWGN}{additive white Gaussian noise}
\newacronym{bpsk}{BPSK}{Binary Phase-Shift Keying}
\newacronym{snr}{SNR}{signal-to-noise ratio}
\newacronym{fer}{FER}{frame-error rate}
\newacronym{sc}{SC}{Successive-Cancellation}
\newacronym{scl}{SCL}{SC List}
\newacronym{scf}{SCF}{SC Flip}
\newacronym{sclf}{SCLF}{Successive-Cancellation List Flip}
\newacronym{dscf}{DSCF}{Dynamic SCF}
\newacronym{pscf}{PSCF}{Partitioned SCF}
\newacronym{cdf}{CDF}{Cumulative Density Function}
\newacronym{pscl}{PSCL}{Partitioned SCL}
\newacronym{psclf}{PSCLF}{Partitioned SCLF}
\newacronym{crc}{CRC}{cyclic-redundancy check}
\newacronym{ca}{CA}{CRC-aided}
\newacronym{llr}{LLR}{log-likelihood ratio}
\newacronym{cc}{CC}{clock cycle}
\newacronym{srm}{SRM}{simplified restart mechanism}
\newacronym{grm}{GRM}{generalized restart mechanism}
\newacronym{ascl}{ASCL}{Adaptive-SCL}
\begin{document}

% Limit number of shown authors, i.e., use "et al." where appropriate
\bstctlcite{IEEEexample:BSTcontrol}

	%\sg{Automorphism-Based Dynamic Frozen-Function Design for RM Codes // with SCL Decoding}
\title{Partitioned Successive-Cancellation List Flip Decoding of Polar Codes}%SCL Decoding of

\author{\IEEEauthorblockN{Charles Pillet\IEEEauthorrefmark{1}, Ilshat Sagitov\IEEEauthorrefmark{1}, Gr\'egoire Domer\IEEEauthorrefmark{2}, and Pascal Giard\IEEEauthorrefmark{1}}%

  \IEEEauthorblockA{\IEEEauthorrefmark{1}Department of Electrical Engineering, \'Ecole de technologie sup\'erieure, Montr\'eal, Qu\'ebec, Canada.\\Email: \{charles.pillet.1,ilshat.sagitov.1\}@ens.etsmtl.ca,  pascal.giard@etsmtl.ca}%

  \IEEEauthorblockA{\IEEEauthorrefmark{2}Department of Electrical Engineering, Enseirb-Matmeca, Bordeaux INP, Bordeaux, France.\\Email: gregoire.domer@enseirb-matmeca.fr}
}
\maketitle

\begin{abstract}
    The recently proposed \gls{sclf} decoding algorithm for polar codes improves the error-correcting performance of state-of-the-art \gls{scl} decoding. 
    However, it comes at the cost of a higher complexity.
    In this paper, we propose the \gls{psclf} decoding algorithm, an algorithm that divides a word in partitions and applies \gls{sclf} decoding to each partition separately.      
    Compared to \gls{sclf}, \gls{psclf} allows early termination but is more susceptible to \gls{crc} collisions. 
    In order to maximize the coding gain, a new partition design tailored to \gls{psclf} is proposed as well as the possibility to support different \gls{crc} lengths.
    Numerical results show that the proposed \gls{psclf} algorithm has an error-correction performance gain of up to 0.15\,dB with respect to \gls{sclf}.
    Moreover, the proposed \gls{crc} structure permits to mitigate the error-correction loss at low \gls{fer} due to \gls{crc} collisions, showing a gain of 0.2\,dB at a \gls{fer} of $\mathbf{10^{-4}}$ with respect to the regular \gls{crc} structure.
    The average execution time of \gls{psclf} is shown to be 1.5 times lower than that of \gls{sclf}, and matches the latency of \gls{scl} at $\text{FER}=\mathbf{4\cdot10^{-3}}$ and lower.    
        
\end{abstract}
\IEEEpeerreviewmaketitle
\glsresetall	%===========================================================================================

\section{Introduction}
    Since the joint invention of polar codes and the asymptotically capacity-achieving \gls{sc} decoding algorithm \cite{ArikanPolarCodes}, progress towards improving the error-correcting performance of \gls{sc} at finite block length has been made by proposing new decoding algorithms of polar codes \cite{SCL,scf_intro}.
    \gls{scl} is the list decoding algorithm based on \gls{sc} \cite{SCL}.
    \gls{scl} tracks in parallel a list of $L$ candidates which improves the error-correcting performance.
    The error-correcting performance of \gls{scl} can further be improved by a concatenation of a \gls{crc} code with the polar code.
    This scheme is referred as \gls{ca}-polar codes and has been chosen as one of the coding scheme in the 5G standard \cite{standard}.

    An alternative decoding algorithm of \gls{ca}-polar codes based on \gls{sc} decoding is \gls{scf} \cite{scf_intro}.
    While the list of candidates was generated in parallel for \gls{scl}, the list of candidates is generated sequentially for \gls{scf} by flipping potential error-prone bits after the first \gls{sc} trial.
    The accuracy of identifying error-prone bits improved in \cite{dyn_scf} while also enabling multi-flipping in additional \gls{sc} trials. 
    \gls{scf} and its variants have a variable execution time, but a throughput and a complexity asymptotically equal to those of \gls{sc}.

    Partitioned polar codes \cite{partition_scl,segmented_scl} are a special type of \gls{ca}-polar codes.
    Partitioned polar codes are segmented into partitions, each of which is protected by its own \gls{crc}.
    %Hence, each \gls{crc} permits to detect errors inside its partition, improving the error detection capability.
    \gls{scl} \cite{partition_scl,segmented_scl} and \gls{scf} \cite{PSCF} support the decoding of partitioned polar codes and are referred to as \gls{pscl} and \gls{pscf}.
    For \gls{pscl}, a coding gain with respect to \gls{scl} has been observed \cite{partition_scl,segmented_scl}.
    Moreover, at equal error-correcting performance, the decoding complexity is reduced \cite{segmented_scl} as well as the memory requirements if the partitions correspond to sub-decoding trees \cite{partition_scl}.
    For \gls{pscf}, the number of flipping trials has been shown to be dividable by a factor $4$ with respect to \gls{scf} for an equivalent error-correcting performance \cite{PSCF}.

    \gls{sclf} is the flip decoding algorithm based on \gls{scl} \cite{first_SCLF}.
    If the first \gls{scl} fails, a list of path-flipping locations is retrieved and \gls{scl} is performed once more and take the $L$ worst paths at the flip location \cite{flip_criteria_real_time,flip_criteria_fixed}.
    Various flip metrics have been proposed to find the path-flipping locations.
    A reliable flipping set based on a heuristic parameter is proposed \cite{metric_with_alpha}.
    Authors in \cite{reduced_complexity_metric} reduce drastically the complexity of this flip metric.
    A dynamic flipping set permitting to adjust the candidate flip locations was proposed in \cite{dyn_sclf}.
    \gls{sclf} supports multiple flip locations \cite{metric_with_alpha,dyn_sclf} but doing this increases the number of decoding attempts, and thus the decoding complexity.
    By combining list and flip decoding strategy, \gls{sclf} returns the state-of-the-art error-correcting performance at the cost of increased complexity and variable execution time.

    In this paper, we propose the \gls{psclf} decoding algorithm. 
    We show that it reduces the decoding complexity of \gls{sclf} and improves its error-correcting performance.
    In order to maximise the coding gain with respect to \gls{sclf}, the partitions are designed according to the main decoder (SCL).
    We show that this approach improves error-correcting performance with respect to other partition design strategies \cite{partition_scl,segmented_scl}.
    Moreover, a \gls{crc} structure is also proposed to reduce \gls{crc} false positives, an issue more frequent in \gls{psclf} than in \gls{sclf}.
    The proposed \gls{crc} structure improves the error-correcting performance at no additional cost.
    Finally, the average execution times of \gls{psclf}  and \gls{sclf} are analyzed and compared to that of \gls{scl} and its adaptive variant \cite{AdaptiveSCL}.    

\section{Preliminaries}

\subsection{Polar Codes}
    A $(N=2^n,K)$ polar code of length $N$ and dimension $K$ is a binary block code based on the polarization effect of the binary kernel $\mathbf{T}_2=\left[\begin{smallmatrix}
        1&0\\1&1
    \end{smallmatrix}\right]$ and of the transformation matrix $\mathbf{T}_N=\mathbf{T}_2^{\otimes n}\in\mathbb{F}_2^{N\times N}$ \cite{ArikanPolarCodes}.
    A $(N,K)$ polar code is fully defined by its information set $\Ical\subseteq[N]\triangleq\{0,\dots,N-1\}$, describing the locations where the message $\mathbf{m}\in\mathbb{F}_2^K$ is inserted in the input vector $\mathbf{u}=(u_0,\dots,u_{N-1})\in\mathbb{F}_2^N$, i.e., $\mathbf{u}_\Ical=\mathbf{m}$.
    The remaining $N-K$ locations, stored in the frozen set $\Fcal=[N]\setminus\Ical$ are set to 0, i.e., $\mathbf{u}_\Fcal=\mathbf{0}$.
    The encoding is performed as $\mathbf{x}=\mathbf{u}\cdot\mathbf{G}_N$, where $\mathbf{x}\in\mathbb{F}_2^N$ is a codeword.

    A $(N,K+C)$ \gls{ca}-polar code is a  $(N,K)$ polar code concatenated with a \gls{crc} code of $C$ bits.
    The \gls{crc} encoder is applied on $\mathbf{m}$ and $C$ \gls{crc} bits are appended to $\mathbf{m}$, defining $\mathbf{m}'\in\mathbb{F}_2^{K+C}$.
    Hence, $\Ical$ is now enlarged by $C$ additional bits.
    As a rule, we have $\Ical=\{i_1,i_2,\dots,i_{K+C}\}$ with $i_1<i_2<\dots<i_{K+C}$.
    In the remaining of the paper, \gls{ca}-polar codes are used and decoded with \gls{sc}-based  \cite{ArikanPolarCodes} algorithms.
    %Hence, \gls{sc} is a sequential decoder.
    \gls{sc} leads to poor error-correcting performance in the finite-length regime but can be improved with more complex \gls{sc}-based decoding algorithm.

	\subsection{SCL Decoding}
    \gls{scl} is the list decoding algorithm also based on \gls{sc}  \cite{SCL}.
    Hence the scheduling is as \gls{sc} and at each information bit $i\in\Ical$, the paths consider both possible values $\{0,1\}$, doubling the size of the list.
    In order to limit the complexity, \gls{scl} only considers $L$ different decoding paths such that path sorting according to a metric is required.
    If $i\in\{i_1,\dots,i_{\log_2(L)}\}$, no path sorting is needed since doubling the number of candidates still generates less than or exactly $L$ candidate paths.
    However, if $i\in\{i_{\log_2(L)+1}, \dots,i_{K+C}\} \triangleq \Icalsort$, the $L$ paths minimizing the path metrics out of the $2L$ candidate paths are selected \cite{SCL_LLR}.
    The path metric $\PM_i[l]$ for index $i\in[N]$ and path $l\in[2L]$ is penalized whenever the bit decision $\hat{u}_i[l]$ of the partial candidate vector $\hat{\mathbf{u}}_0^i[l]\in\mathbb{F}_2^{i+1}$ does not correspond to the hard decision of the \gls{llr} $\Lij{0}{i}[l]$ in the $i^{th}$ leaf of path $l$.
    \gls{scl} returns $L$ candidates for $\mathbf{u}$, noted $\hat{\mathbf{u}}[l]$ with $0\leq l\leq L-1$.
    The candidate passing the \gls{crc} with the lowest path metric is chosen as decoder output $\mathbf{\hat{u}}$ which improves the error-correction performance.

    \gls{ascl} \cite{ASCL} is an iterative decoding algorithm successively increasing the list size $L$ up to a maximum and stops as soon as the CRC code is checked.
    The complexity of \gls{ascl} converges to \gls{sc} at low \gls{fer}, but the time complexity increases and depends from the channel condition.
    \subsection{SCF Decoding}
    \gls{scf} decoding performs up to $T_{\max}$ SC decoding trials. 
    A \gls{crc} check is performed at the end of each trial. 
    If the first trial fails, a set of flipping candidates $\beta$ is generated with $|\beta|=T_{\max}-1$. 
    The $t^{th}$ additional trial performs SC except at the flipping location $\beta_t$ where the reverse decision on $\hat{u}_{\beta_{t}}$ is taken.
    The decoding latency of \gls{scf} is $\latSCF =T_{\max}\times \latSC$ while its average execution time $\avgSCF$ corresponds to
    \begin{align}
        \avgSCF = \avgt \times \latSC \leq \latSCF,\label{eq:avgSCF}
    \end{align}
    where $\avgt\leq T_{\max}$ is the average number of decoding trials.
    
    \gls{pscf} \cite{PSCF} is a flip decoder allowing to decode \gls{ca}-polar codes with multiple \glspl{crc} distributed in the codeword.
    A partitioned polar code is partly described by its  set $\mu$ corresponding to the last indices of all partitions.
    The size $P=|\mu|$ describes the number of partitions.

    %Recently \cite{SRM_WCNC,GRM_TSP}, a restart mechanism has been proposed to reduce the average execution times of SCF-based decoders.
    %The mechanism applies conditional restart based on the flip location.
    %In the \gls{grm}, the restart location corresponds to the smallest location where a flip is applied leading to the largest average execution time gain.
    %However, the \gls{grm} requires a complex implementation such that the \gls{llrm} was proposed.

    \subsection{SCLF Decoding}
    \gls{sclf} algorithm \cite{first_SCLF} is the flip decoding algorithm of polar codes with \gls{scl} as the core decoder. 
    The best flipping strategy has been proposed in \cite{flip_criteria_fixed,flip_criteria_real_time}.
    On a flip location $i\in\Icalsort$, the $L$ worst paths are selected instead of the best $L$ paths \cite{SCL}.

    If we note the flip metric $\FM_i^\alpha$ at index $i\in\Icalsort$ and normalized with $\alpha\geq 1$, a factor mitigating the impact of propagation of errors. 
    A reliable flip metric was proposed in \cite{metric_with_alpha}, later simplified for hardware purpose in \cite{reduced_complexity_metric} as
    \begin{align}\label{eq:reduced_complexity_metric}
        \FM_i^\alpha=-\PM_i[0]+\alpha\PM_i[L],
    \end{align}
    where the path metrics are sorted from most to least reliable.
    Since a lower complexity \gls{sclf} variant is targeted, \eqref{eq:reduced_complexity_metric} is selected.
    The flipping set $\mathcal{B}_{\text{flip}}\subset\Icalsort$, storing the indices where the flips are performed, is designed such that $\mathcal{B}_{\text{flip}}(t)$ is the index $i\in\Icalsort$ with the $t^{th}$ lowest flip metric \cite{metric_with_alpha}.
    
 %   \begin{align}
  %          \mathcal{B}_{\text{flip}}=\left\{i\,\, |\,\,\overunderset{T}{i\in\Ical_{\text{sort}}}{\mathrm{argmin}}\left(\FM_\alpha^i\right)\right\}
 %       \end{align}
 %   where $\overunderset{T}{i\in\Ical_{\text{sort}}}{\mathrm{argmin}}$ denotes the function $\mathrm{argmin}$ on the $T$ minimum values.
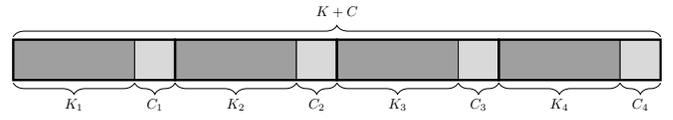
\begin{figure}[t]
    \centering
    \resizebox{.99\columnwidth}{!}{\usetikzlibrary{decorations.pathreplacing}

\begin{tikzpicture}
    \newcommand{\height}{1}
    % Rectangle principal de 16 cm de Largeur et 4 cm de hauteur
    \draw (0,0) rectangle (16,\height);
    
    % Divisions : 3cm, 1cm, 3cm, 1cm, 3cm, 1cm, 3cm, 1cm
    \draw[fill=gray!75] (0,0) rectangle (3,\height);
    \draw[fill=gray!30] (3,0) rectangle (4,\height);
    
    \draw[fill=gray!75] (4,0) rectangle (7,\height);
    \draw[fill=gray!30] (7,0) rectangle (8,\height);
    \draw[fill=gray!75] (8,0) rectangle (11,\height);
    \draw[fill=gray!30] (11,0) rectangle (12,\height);
    \draw[fill=gray!75] (12,0) rectangle (15,\height);
    \draw[fill=gray!30] (15,0) rectangle (16,\height);

    % delimitation of partition
    \draw[ultra thick] (0,0) rectangle (4,\height);
    \draw[ultra thick] (4,0) rectangle (8,\height);
    \draw[ultra thick] (8,0) rectangle (12,\height);
    \draw[ultra thick] (12,0) rectangle (16,\height);
    % Accolades
    \draw [decorate,decoration={brace,amplitude=8pt,mirror,raise=2pt}] (0,0) -- (3,0) node[midway,below=10pt] {$\Large K_1$};
    \draw [decorate,decoration={brace,amplitude=8pt,mirror,raise=2pt}] (3,0) -- (4,0) node[midway,below=10pt] {$\Large C_1$};
    \draw [decorate,decoration={brace,amplitude=8pt,mirror,raise=2pt}] (4,0) -- (7,0) node[midway,below=10pt] {$\Large K_2$};
    \draw [decorate,decoration={brace,amplitude=8pt,mirror,raise=2pt}] (7,0) -- (8,0) node[midway,below=10pt] {$\Large C_2$};
    \draw [decorate,decoration={brace,amplitude=8pt,mirror,raise=2pt}] (8,0) -- (11,0) node[midway,below=10pt] {$\Large K_3$};
    \draw [decorate,decoration={brace,amplitude=8pt,mirror,raise=2pt}] (11,0) -- (12,0) node[midway,below=10pt] {$\Large C_3$};
    \draw [decorate,decoration={brace,amplitude=8pt,mirror,raise=2pt}] (12,0) -- (15,0) node[midway,below=10pt] {$\Large K_4$};
    \draw [decorate,decoration={brace,amplitude=8pt,mirror,raise=2pt}] (15,0) -- (16,0) node[midway,below=10pt] {$\Large C_4$};
    \draw [decorate,decoration={brace,amplitude=8pt,raise=2pt}] (0,\height) -- (16,\height) node[midway,above=12pt] {$\Large K+C$};
    
\end{tikzpicture}}
    \caption{Message $\mathbf{m}'\in\mathbb{F}_2^{K+C}$ allocated in $\Ical$ for a partitioned polar code with $P=4$ partitions. Dark gray corresponds to the message $\mathbf{m}$ and light gray corresponds to the \gls{crc} bits.}
    \label{fig:message_partitione}
\end{figure}
\section{Partitioned SCLF decoder}
Both \gls{scl} and \gls{scf} have been adapted to decode partitioned polar codes. The \gls{pscl} and \gls{pscf} algorithms  were shown to have an improved error-correction performance and a reduced decoding complexity over their respective counterparts.
\gls{sclf} is a more complex algorithm than \gls{scf} or \gls{scl}, hence we propose the \gls{psclf} algorithm and study the complexity reduction.
\subsection{Partitioned Polar Codes}
A partitioned polar code is divided into $P>1$ partitions. 
Each partition should contain $K_p\geq 1$ information bits and is concatenated with its own \gls{crc} code of size $C_p$.
The number of information bits and \gls{crc} bits verify $K=\sum_{p=1}^P K_p$ and $C=\sum_{p=1}^P C_p$.
Thus, the partitioned polar code is as well defined with $|\Fcal|=N-K-C$ and $|\Ical|=K+C$.
\autoref{fig:message_partitione} depicts the message $\mathbf{m}'\in\mathbb{F}_2^{K+C}$ for $P=4$ partitions.
The vector $\mathbf{m}'$ is then allocated in the positions stated in $\Ical$, retrieving $\mathbf{u}$ and the polar encoding is performed to retrieve the codeword $\mathbf{x}$.

In the following, the number of non-frozen bits in the $p^{\text{th}}$ partition is noted $s_p=K_p+C_p$ while the cumulative number of non-frozen bits is noted $S_p=\sum_{i=1}^{p}s_p$ where $S_1\triangleq s_1$ and $S_P\triangleq K+C$.
Inside a codeword, the last indices of each partition are stored in $\mu=\{\mu_1,\dots, N-1\}$ and $\mu_p\triangleq\Ical\left(S_p\right)=i_{S_p}$.
%Next, the $p^{\text{th}}$ partition can be defined by a 4-tuple $\left(p,K_p,C_p,\mu_p\right)$.

A elementary design of partitions $\mu$ is to uniformly distribute the partitions based on the information set $\Ical$ \cite{segmented_scl}.
Namely, it corresponds to the case where $s_p=K_p+C_p=\frac{K+C}{P}$ for any $p$.
Such design of $\mu$ will be denoted $\mu_{\text{eq}}$.
For a code $(1024,512+32)$ and $P=4$, each partition contains $\frac{K+C}{P}=136$ non-frozen bits.
Another simple design of the partition $\mu$ is to consider partitions as sub-decoding trees \cite{partition_scl}. 
For $P$ partitions, each partition is of length $\frac{N}{P}$.
However, this approach may lead to partitions without information bits for low-rate codes. Therefore, this approach is not considered.

\subsection{Partition Design Based on SCL}
%Given a partitioned polar code, the design of $\mu$ has an impact on the decoding performance.
In this paper, partitioned polar codes are decoded with \gls{psclf}, a flip decoder based on \gls{scl}.
%Since \gls{scl} is used as main decoder, 
Hence, we propose to design the partitions, i.e., the set $\mu$ according to the decoding behaviour of \gls{scl}.
Next, we denote by $X$ the random variable describing the first error in \gls{scl} for a polar code defined by $\Ical$.
The first error in \gls{scl} corresponds to the first index $i\in[N]$ where $\forall l\in[0,L-1], \mathbf{\hat{u}}_0^i[l]\neq\mathbf{u}_0^i$, i.e., all paths have diverged.
%The definition is valid for any list size $L$, and also for $L=1$, corresponding to SC decoder.
If $i\in\Fcal$, no paths duplication is performed such that the first error cannot occur. 
It cannot also occur when all decoding paths are covered, namely if $i\in\{i_1,\dots,i_{\log_2(L)}\}$.
Hence, the events of $X$ are only possible in $\{i_{\log_2(L)+1},\dots,i_{K+C}\}\triangleq\Icalsort$.
The \gls{cdf} of $X$ is denoted $F(k)$, and is defined as 
\begin{align}
    F(k) = \mathbb{P}(X \leq k)=\sum_{i=0}^k \mathbb{P}(X=i)\,,
\end{align}
where $\mathbb{P}(X=i)=0$ if $i\in\Fcal$ or $i\in\{i_1,\dots,i_{\log_2(L)}\}$.

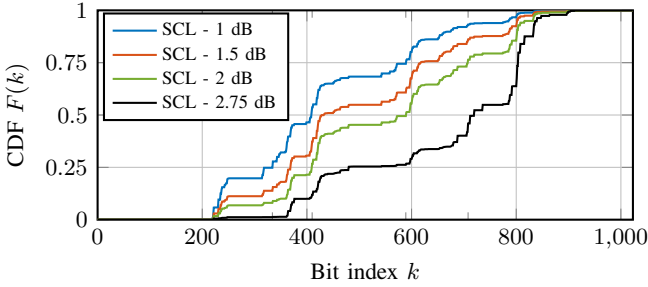
\begin{figure}
    \centering
    \vspace{2pt}
    \resizebox{.99\columnwidth}{!}{\begin{tikzpicture}
  \pgfplotsset{
    label style = {font=\fontsize{9pt}{7.2}\selectfont},
    tick label style = {font=\fontsize{9pt}{7.2}\selectfont}
  }
   
   \begin{axis}[%
    width=\columnwidth,
    height=0.5\columnwidth,
    xmin=0, xmax=1023,
    minor xtick={334,410,589,708},
    xlabel={Bit index $k$},
    xlabel style={yshift=0.4em},
    ymin=0, ymax=1,
    ytick={0,0.25,0.5,0.75,1},
    ylabel style={yshift=-0.4em},
    ylabel={CDF $F(k)$},
    xlabel style={yshift=-0.2em},
    yminorticks, xmajorgrids,
    ymajorgrids, yminorgrids,
    legend style={at={(0.01,0.99)},anchor=north west},
    legend style={nodes={scale=0.8}, font=\small},
     legend style={legend columns=1},
     legend cell align={left},
    mark size=1.6pt, line width=0.8pt, mark options=solid,
    ] 

 % \addplot [mark=triangle, color=MyBlue, mark repeat=50, mark phase=50]
 % table[x=bfidx,y=CyclesSavedP16]{data/Saved_cycles_N1024_P.tex};
 % \addlegendentry{Saved \gls{cc} \eqref{eq:l_p_saved_calc} $P=16$} 

 \addplot [solid, color=matlab1, mark phase=0]
 table[x=xdata,y=ydata]{data/CDF_L4_1024_512_32_1dB.data};
 \addlegendentry{\gls{scl} - 1 dB}

  \addplot [solid, color=matlab2, mark phase=0]
 table[x=xdata,y=ydata]{data/CDF_L4_1024_512_32_1.5dB.data};
 \addlegendentry{\gls{scl} - 1.5 dB}

   \addplot [solid, color=matlab5, mark phase=0]
 table[x=xdata,y=ydata]{data/CDF_L4_1024_512_32.data};
 \addlegendentry{\gls{scl} - 2 dB}
    \addplot [solid, color=black, mark phase=0]
 table[x=xdata,y=ydata]{data/CDF_L4_1024_512_32_2.75dB_new.data};
 \addlegendentry{\gls{scl} - 2.75 dB}
\end{axis}    
\end{tikzpicture}%
}
    \caption{\gls{cdf} $F(k)$ of $(1024,512+32)$ polar code for $L=4$ and $\frac{E_b}{N_0}=\{1,\,1.5,\, 2,\,2.75\}$ dB.}
    \label{fig:CDF_SNR}
\end{figure}
\autoref{fig:CDF_SNR} depicts $F(k)$ for $L=4$ of a $(1024,512+32)$ polar code at $\frac{E_b}{N_0}=\{1,\, 1.5,\, 2,\,2.75\}$\,dB.
As seen in \autoref{fig:CDF_SNR}, the probability that the first error occurs earlier in $\hat{u}$ increases with the noise.
For instance, the location delimiting half of the first errors, i.e., $F(k)=0.5$ is reached at $k=\{409,433,590,720\}$ for $\frac{E_b}{N_0}=\{1,\, 1.5,\, 2,\,2.75\}$\,dB, respectively.
%Overall, the shape of all \gls{cdf} is identical but with a vertical shift when the noise is higher.

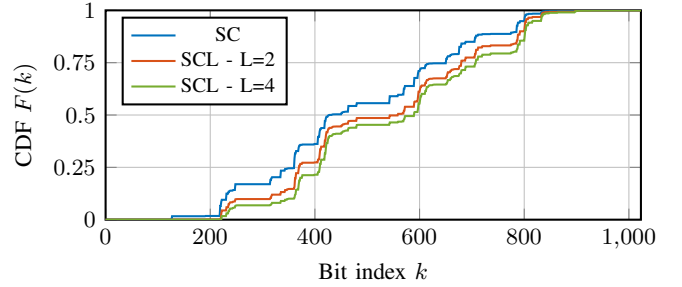
\begin{figure}[t]
    \centering
    \vspace{2pt}
    \resizebox{.99\columnwidth}{!}{\begin{tikzpicture}
  \pgfplotsset{
    label style = {font=\fontsize{9pt}{7.2}\selectfont},
    tick label style = {font=\fontsize{9pt}{7.2}\selectfont}
  }
   
   \begin{axis}[%
    width=\columnwidth,
    height=0.5\columnwidth,
    xmin=0, xmax=1023,
    xlabel={Bit index $k$},
    xlabel style={yshift=0.4em},
    ymin=0, ymax=1,
    ytick={0,0.25,0.5,0.75,1},
    ylabel style={yshift=-0.4em},
    ylabel={CDF $F(k)$},
    xlabel style={yshift=-0.2em},
    yminorticks, xmajorgrids,
    ymajorgrids, yminorgrids,
    legend pos=north west,
    legend style={nodes={scale=0.8}},
     legend style={legend columns=1},
    mark size=1.6pt, line width=0.8pt, mark options=solid,
    ] 

 % \addplot [mark=triangle, color=MyBlue, mark repeat=50, mark phase=50]
 % table[x=bfidx,y=CyclesSavedP16]{data/Saved_cycles_N1024_P.tex};
 % \addlegendentry{Saved \gls{cc} \eqref{eq:l_p_saved_calc} $P=16$} 

 \addplot [solid, color=matlab1, mark phase=0]
 table[x=xdata,y=ydata]{data/CDF_1_1024_512_32.data};
 \addlegendentry{\gls{sc}}

  \addplot [solid, color=matlab2, mark phase=0]
 table[x=xdata,y=ydata]{data/CDF_L2_1024_512_32.data};
 \addlegendentry{\gls{scl} - L=2}

   \addplot [solid, color=matlab5, mark phase=0]
 table[x=xdata,y=ydata]{data/CDF_L4_1024_512_32.data};
 \addlegendentry{\gls{scl} - L=4}
\end{axis}    
\end{tikzpicture}%
}
    \caption{\gls{cdf} $F(k)$ of $(1024,512+32)$ polar code for various list sizes $L$ at $\frac{E_b}{N_0}=2$ dB.}
    \label{fig:CDF_L}
\end{figure}
\autoref{fig:CDF_L} depicts $F(k)$ of a $(1024,512+32)$ polar code for various list sizes $L$, all at a $\frac{E_b}{N_0}=2$\,dB. % and at least $20000$ first errors have been computed to have an accurate value of $P(X=i)$.
As seen in \autoref{fig:CDF_L}, as the list size $L$ grows, the probability to have a first error early in the frame decreases.
It is explained by the shift to the end of the first index of $\Icalsort$ and the improved error capability of \gls{scl} with a larger list size.
The location delimiting a quarter of the first errors, i.e., $F(k)= 0.25$ is reached at $k=\{361, 370 , 410\}$ for $L=\{1,2,4\}$, respectively.

The proposed design of $\mu$ is to uniformly distribute the partitions according to the \gls{cdf} $F(k)$.
Namely, $\forall p\in\{1,\dots,P\}, \mu_p$ verifies
\begin{align}
    F(\mu_p-1)<\frac{p}{P}\leq F(\mu_p).
\end{align}
A similar design for \gls{pscf} was carried out by using the decoding behaviour of \gls{sc} \cite{PSCF}.
Since the proposed \gls{psclf} algorithm uses \gls{scl} as its core decoder, the \glspl{cdf} are based on the decoding behaviour of \gls{scl}.
Given \autoref{fig:CDF_SNR} and for a simulation using \gls{scl} with $L=4$, the two sets $\mu$ are
\begin{align}
    \mu&=\{335,409,589,1023\},\label{eq:mu_1dB}\\
    %\mu_{\text{sim}}=\{361,428,651,1023\},\\
    %\mu_{\text{sim}-2}=\{370,559,680,1023\},\\
    \mu&=\{410,590,708,1023\},\label{eq:mu_2dB}    
\end{align}
for a design at $1$ and $2$ dB.
The number of non-frozen bits in each partition is based on $\mu$ and the information set $\Ical$ of the polar code.
For these particular examples, the number of non-frozen bits $\mathcal{K}=\{s_1,s_2,s_3,s_4\}$ inside each partition is $\mathcal{K}=\{28,32,100,384\}$ if $\mu$ \eqref{eq:mu_1dB} and $\mathcal{K}=\{61,100,84,299\}$ if $\mu$ \eqref{eq:mu_2dB}.
All partitions are protected by a \gls{crc} code and given  the sets $\mathcal{K}$, the number of non-frozen bits fluctuates heavily with the partitions.
For both $\mu$, $s_1$ is  small while $s_P$ is large.
Next, we discuss the \gls{crc} structure of partitioned polar codes.
\subsection{CRC Structure of the Partitioned Polar Codes}
%Based on \eqref{eq:nb_info_bit_partition}, the number of non-frozen bits is unequally distributed with our proposed design of set $\mu$.
%Hence, the number of \gls{crc} bits applied on each partition could also vary and match this pattern.
With the exception of \cite{CRC_PSCL_virtual}, partitioned decoding of polar codes is usually carried out with a fixed number of \gls{crc} bits per partitions. Defining the set of \gls{crc}-code size per partition as $\mathcal{C}=\{C_1,\dots,C_P\}$, most works have $C_1=\dots=C_P$.
In \cite{CRC_PSCL_virtual}, the \gls{crc} structure was designed according to the capacity of the sub-channels given a set $\mu$. 
As a consequence, this method designs the \gls{crc} structure according to the length of the partitions as well as the reliability of the polar code.
However, the design of $\mu$ was not discussed and $\mu=\{255,511,767,1023\}$ was chosen with $\mathcal{K}=\{20,123,156,245\}$ non-frozen bits in each partition.
The resulting \gls{crc} structure is $\mathcal{C}_1=\{3,11,10,8\}$ \cite{CRC_PSCL_virtual}.

\begin{table}[t!]
    \vspace{10pt}
    \centering
    \begin{tabular}{|c|>{\centering}p{1.2cm}|>{\centering}p{1.2cm}|>{\centering}p{1.2cm}|c|}
        \cline{2-5}
         \multicolumn{1}{c|}{} &\multicolumn{4}{c|}{Partition $p$}  \\
        \cline{2-5} 
         \multicolumn{1}{c|}{} &1&2&3&4  \\
         \hline 
         1&0.53&0.23&0.17&0.07\\
         1.5&0.37&0.25&0.24&0.14\\
         2&0.25&0.25&0.25&0.25\\
         2.75&0.09&0.14&0.21&0.56\\
         \hline
         $\nicefrac{E_b}{N_0}$ dB&\multicolumn{4}{c|}{Probability $\mathbb{P}(e_p)$ that the first error occurs in partition $p$.}\\
         \hline
    \end{tabular}
    \caption{Probabilities that the first error occurs in each partition, the set $\mu$ is designed at $2$ dB \eqref{eq:mu_2dB}.} %$\mathbb{P}\left(\mu_{p-1}<X\leq\mu_{p}\right)$
    \label{tab:prob_error}
\end{table}
Our proposed \gls{crc} structure $\mathcal{C}$ is designed according to the probability of error in each partition and the decoding behaviour of \gls{scl}.
In a partition, an error is either caused by having no paths passing the \gls{crc} or having one path passing the \gls{crc} but being a false positive, i.e, a \gls{crc} collision.
The former may end up getting corrected by a decoding-path flip.
The latter, however, cannot be corrected by \gls{psclf} since only wrong paths will end up being forwarded to the next partition.
Hence, \gls{crc} collisions should be reduced and the proposed \gls{crc} structure does so by taking the probability of errors in each partition into account.

Given \autoref{fig:CDF_SNR}, at low $\frac{E_b}{N_0}$ values, a decoding error is mostly due to an error happening in the earlier indices of a codeword  while at high $\frac{E_b}{N_0}$ values a decoding error is mostly occurring in the latter indices.
\autoref{tab:prob_error} shows the probability that the first error occurs for each partition, the set $\mu$ is designed at $2$ dB \eqref{eq:mu_2dB}.
For a given partition $p$ and $\frac{E_b}{N_0}$, the probability $\mathbb{P}(e_p)$ that the first error occurs can be retrieved from \autoref{fig:CDF_SNR}, i.e.,
\begin{align}
    \mathbb{P}(e_p)=F(\mu_p)-F(\mu_{p-1}+1).
\end{align}
The number of \gls{crc} bits are allocated according to the target \gls{snr}. 
If the design is performed at $\frac{E_b}{N_0}=1$ dB, more \gls{crc} bits should be allocated in the first partition since $P(e_1)=53$\%.
If the design is performed $\frac{E_b}{N_0}=2.75$ dB, more \gls{crc} bits should be allocated towards the last partition since $p(e_4)=56$\%.

%Next, the event of a collision happening at partition $p$ is noted $E_p$. 
%The probability of a collision with respect to the total number of errors is noted $\mathbb{P}(E_P)$.
%Results are available in Section~\ref{subsec:CRC_sim}.
\subsection{Description of \gls{psclf} Decoding}
%Next, the description of \gls{psclf} is carried out.
\Gls{psclf} decoding requires knowledge of $\mathcal{C}$ and $\mu$.
On each partition, up to $T_{\max}$ \gls{scl} trials are performed.
After reaching $\mu_1$ for the first time, the flipping set $\beta$ is computed only when none of the paths satisfy the \gls{crc} code.
If $T_{\max}$ trials are performed and still no paths satisfy the \gls{crc}, decoding failure is raised.
If at a trial $1\leq t \leq T_{\max}$, the \gls{crc} is checked for at least one path, the decoding of the next partition begins.

For the sake of simplicity, the \emph{check and keep} method, as described in \cite{5G_DCA_WCNC}, is used when the \gls{scl} decoder faces \gls{crc} bits inside a codeword.
In this method, when performing a \gls{crc} check, both paths that satisfy the \gls{crc} and those that do not are kept and forwarded to the next partition.
This method corresponds to the original \gls{scl} algorithm since no changes are made to the algorithm, however, it suffers from a small performance degradation compared to other methods requiring more data control \cite{5G_DCA_WCNC}.
%Check and Remove

Decoding is successful if all $P$ \gls{crc} codes corresponding to each partitions are satisfied. 
\Gls{psclf} naturally implements \emph{early termination} as decoding stops before the end of a frame whenever a decoding failure happens at a partition $1\leq p<P$.

\subsection{Average Execution Time of \gls{psclf}}\label{subsec:exec_psclf}
%Next, the average execution time of \gls{psclf} is analysed.
In \cite{PSCF}, the complexity of \gls{pscf} was linked to the \emph{normalized average computational complexity}. 
However, no equations were given to compute it.
In the following, we provide detailed equations to compute the average execution time of both \gls{pscf} and \gls{psclf}.
Since the partitions are of different lengths, the impact of flipping fluctuates from partition to partition.
The partial latency $\latSC\left(i\right)$ required by the semi-parallel SC decoder to decode until the bit $i\in[N]$ is derived from \cite{GRM_TSP}
\begin{align}
 \label{eq:l_p_saved_calc}
 \latSC\left(i\right) &= \sum_{s=0}^{n-1}\left\lceil \frac{2^s}{\varphi}\right  \rceil +\sum_{s=0}^{n-1} \left( \left \lceil \frac{2^s}{\varphi}\right  \rceil \times \left \lfloor \frac{i}{2^s} \right \rfloor \right),
\end{align}
where $\varphi$ is the number of processing elements.
If $i=N-1$ and corresponds to the end of the decoding, the latency of \gls{sc} is retrieved, i.e., $\latSC=\left(2N + \frac{N}{\varphi}\cdot \log_2{\left(\frac{N}{4\varphi}\right)}\right)$ \cite{semi_par_sc}.
%By drawing a parallel with $\avgSCF$ \eqref{eq:avgSCF}, the average execution time $\avgPSCF$ of the \gls{pscf} decoder corresponds to the summation of the average execution time spent in each partition.
%Hence, \eqref{eq:l_p_saved_calc} is derived to compute $\avgPSCF$, namely
%\begin{align}
    %\avgPSCF = \sum_{p=1}^{P} \overline{t_p}\times\left(\latSC(\mu_p)-\latSC(\mu_{p-1})\right),\label{eq:avgPSCF}
%\end{align}
%where $1\leq\overline{t_p}\leq T_{\max}$ is the average number of \gls{sc} trials to decode the $p^{\text{th}}$ partition and by convention $\latSC(\mu_{0}=0)=0$. 
%The decoding latency $\latPSCF$ is retrieved if $\forall p\in[1,P], \overline{t_p}=T_{\max}$, we have $\latPSCF=\latSCF$ by \eqref{eq:avgPSCF}.

%The latency of \gls{scl} $\latSCL$ depends on the number of sorting operations and is $\latSCL=\latSCL(N-1)=\latSC+|\Ical|$ \cite{PolarBear}.
The latency of \gls{scl} is given in \cite{PolarBear} as $\latSCL=\latSCL(N-1)=\latSC+|\Ical|$.
Similarly to \gls{scf}'s \eqref{eq:avgSCF}, the average execution time of \gls{sclf} is $\avgSCLF=\latSCL\times\overline{t}$ where $\overline{t}$ is the average number of \gls{scl} trials per frame.
The partial \gls{scl} latency $\latSCL\left(i\right)$ depends on the number of sorting operations required up to index $i$. 
Given that $\mathcal{K}=\{s_1,\dots,s_P\}$ is the set storing the number of non-frozen bits in each partition, the \gls{scl} partial latency $\latSCL\left(\mu_p\right)$ to decode the first $p$ partitions is
\begin{align}
 \label{eq:latSCL}
 \latSCL\left(\mu_p\right) &= \sum_{m=1}^{p}s_m + \latSC(\mu_p),
\end{align}
where the left term represents the number of sorting operations by the end of the $p^{\text{th}}$ partition.
To compute the average execution of \gls{psclf}, early-termination has to be taken into account.
Next, the probability that decoding is performed in partition $p$ is noted $\mathbb{P}(T_p)$. 
The probability is linked to early-termination since it corresponds to the probability that decoding has not stopped early in any of the previous $p-1$ partitions.
For the first partition, the probability is $\mathbb{P}(T_1)=1$.
The average execution time of \gls{psclf} corresponds to the summation of the average time spent in each partition mitigated by the probability $\mathbb{P}(T_p)$ of decoding it:
\begin{align}
    \avgPSCLF = \sum_{p=1}^{P} \mathbb{P}(T_p)\,\overline{t_p}\left(\latSCL(\mu_p)-\latSCL(\mu_{p-1})\right)\label{eq:avgPSCLF},
\end{align}
where $1\leq\overline{t_p}\leq T_{\max}$ is the average number of \gls{scl} trials to decode the $p^{\text{th}}$ partition and by convention $\latSCL(\mu_{0}=0)=0$. 
%where $\latSCL(\mu_{p})$ denotes the latency required by \gls{scl} to decode until index $\mu_p$. 

%The latency of \gls{scl} correspond to the latency of \gls{sc} incremented by $K$ and corresponding to the $K$ sorting operations on indices in $\Ical$.
\section{Simulation results}\label{sec:sim_results}
All simulations are performed over the \gls{awgn} channel using the \gls{bpsk} modulation.
First, code parameters used are $N=1024$, $K=512$, and $C=32$.
Then, $N=\{512,2048\}$ and $K=\{256,1024\}$ are used to confirm the results.
The information set is designed at $\frac{E_b}{N_0}=2$ dB (1dB for $N=2048$) while $L=4$ is used.
With $P=1$, performances of SCL-4 ($\mathbin{
\tikz[baseline]{\draw[color=black,dashed,-,thick] (0pt,.5ex) -- (2ex,.5ex) -- (4ex,.5ex);}}$), SCLF ($\mathbin{
\tikz[baseline]{\draw[color=black,solid,thick] (0pt,.5ex) -- (2ex,.5ex) -- (4ex,.5ex);
\node[diamond,draw,solid, color=black,thick, aspect=0.5, text width=4ex, inner sep=-5pt] (d) at (2ex,.5ex) {};
}}$), and \gls{ascl}-L with $L>4$ ($\mathbin{
\tikz[baseline]{ \draw[color=black,dashed,-,thick] (0pt,.5ex) -- (2ex,.5ex) -- (4ex,.5ex);
\draw[color=black,-,thick] (1.5ex,.0ex) rectangle (2.5ex,1ex);}}$) are shown for references.
The latter will match the performance of our proposed \gls{psclf} algorithm.
We set $T_{\max}=15$ and the approximation function \eqref{eq:reduced_complexity_metric} with $\alpha=1.2$ is used \cite{metric_with_alpha,reduced_complexity_metric}.
%the parameter $\alpha$ \eqref{eq:equation_alpha_exp}-\eqref{eq:reduced_complexity_metric}  is optimized and set to $\alpha=1.2$ according to~\cite{metric_with_alpha}. 
\subsection{Impact of Partition Designs}
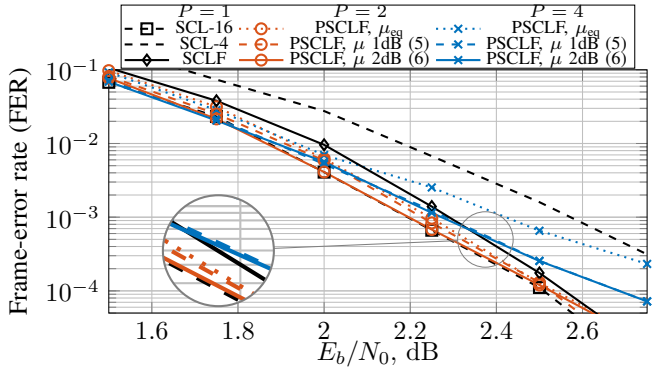
\begin{figure}[t!]
    \centering
    \vspace{4pt}
    \resizebox{.99\columnwidth}{!}{\usetikzlibrary{spy}
\begin{tikzpicture}[spy using outlines={circle, magnification=2, connect spies}]
  \pgfplotsset{
    label style = {font=\fontsize{10pt}{8.2}\selectfont},
    tick label style = {font=\fontsize{10pt}{8.2}\selectfont}
  }

   \begin{semilogyaxis}[
    width=\columnwidth,
    height=0.55\columnwidth,
    xmin=1.5, xmax=2.75,
    xlabel={$E_b/N_0,\,\mathrm{dB}$},
    xlabel style={yshift=0.8em},
    ymin=5e-5, ymax=0.1,
    ylabel style={yshift=-0.1em},
    ylabel={Frame-error rate (FER)},
    yminorticks, xmajorgrids,
    ymajorgrids, yminorgrids,
    legend style={at={(0.5,1.04)},anchor=base},
    %column font, row sep
    legend style={legend columns=3, font=\scriptsize, row sep=-1.5mm},
    legend style={fill=white, fill opacity=1, draw opacity=1,text opacity=1}, % for future use maybe ? %opacity of filling/border and inside text
    legend style={inner xsep=1pt, inner ysep=-2pt}, % TIGHTER
    mark size=1.6pt, mark options=solid,
    ]   
    \addlegendimage{empty legend}
    \addlegendentry{$P=1$}
    \addlegendimage{empty legend}
    \addlegendentry{$P=2$}
    \addlegendimage{empty legend}
    \addlegendentry{$P=4$}
   \addplot[color=black, dashed, mark=square, line width=0.8pt, mark size=2.1pt]
    table[x=Eb,y=FER]{data/Adaptive_SCL16_1k.data};
    \addlegendentry{SCL-16}      
   
       \addplot[color=matlab2, mark=o, dotted, line width=0.8pt, mark size=2.1pt]
     table[x=xdata,y=ydata]{data/data_FER/P2_mu_eq.data};    
    \addlegendentry{PSCLF, $\mu_\text{eq}$}    

         \addplot[color=matlab1, mark=x, dotted, line width=0.8pt, mark size=2.1pt]
     table[x=xdata,y=ydata]{data/data_FER/mu_eq.data};    
    \addlegendentry{PSCLF, $\mu_\text{eq}$}

\addplot[color=black, dashed, line width=0.8pt, mark size=2.1pt]
    table[x=xdata,y=ydata]{data/data_FER/scl.data};    
    \addlegendentry{SCL-4}

    \addplot[color=matlab2, mark=o, dashed, line width=0.8pt, mark size=2.1pt]
table[x=xdata,y=ydata]{data/data_FER/P2_mu_sim_1dB.data}; 
    \addlegendentry{PSCLF, $\mu$  1dB \eqref{eq:mu_1dB}} 
    
  \addplot[color=matlab1, mark=x, dashed, line width=0.8pt, mark size=2.1pt]
table[x=xdata,y=ydata]{data/data_FER/mu_sim_1dB.data}; 
    \addlegendentry{PSCLF, $\mu$ 1dB \eqref{eq:mu_1dB}}

       \addplot[color=black, mark=diamond, line width=0.8pt, mark size=2.1pt]
table[x=xdata,y=ydata]{data/data_FER/sclf.data}; 
    \addlegendentry{SCLF}  
\addplot[color=matlab2, solid,mark=o, line width=0.8pt, mark size=2.1pt]
   table[x=xdata,y=ydata]{data/data_FER/P2_mu_sim_2dB.data}; 
    \addlegendentry{PSCLF, $\mu$ 2dB \eqref{eq:mu_2dB}}
\addplot[color=matlab1, solid,mark=x, line width=0.8pt, mark size=2.1pt]
   table[x=xdata,y=ydata]{data/data_FER/mu_sim_2dB.data}; 
    \addlegendentry{PSCLF, $\mu$ 2dB \eqref{eq:mu_2dB}}
   
    \coordinate (spypoint) at (axis cs:2.375,0.0005); %xy to spy
    \coordinate (magnifyglass) at (axis cs:1.755,0.000352); %sy to zoom
  \end{semilogyaxis}

  \spy [gray, size=1.5cm] on (spypoint) in node[fill=white] at (magnifyglass);

\end{tikzpicture}%

% \addplot[color=black,dashed, mark = x,  line width=0.8pt, mark size=2.1pt]
 %    table[row sep=crcr]{%
 %    1.0000    7.4700e-01  \\
 %    1.1250     6.1600e-01 \\
 %    1.2500    5.1500e-01 \\
 %    1.3750     3.8300e-01 \\
 %    1.5000      3.0300e-01 \\
 %    1.6250     1.8900e-01 \\
 %    1.7500     1.4600e-01\\
 %    1.8750     7.7399e-02  \\
 %    2.0000    4.5086e-02  \\
 %    2.1250    3.5984e-02 \\
 %    2.2500     1.4162e-02\\
 %    2.3750   9.7314e-03 \\
 %    2.5000   4.6904e-03  \\
 %    2.6250   2.1277e-03  \\  
 %    2.75  1.1289e-03  \\
 %    2.875  4.0902e-04  \\
 %    3.0  1.7054e-04 \\
 %    };   
 %    \addlegendentry{P=1 orig. metr. matlab} 

    % \addplot[color=red,dashed,  mark=x, line width=0.8pt, mark size=2.1pt]
    % table[row sep=crcr]{%   
    % 1.0000e+00 7.4200e-01 \\
    % 1.1250e+00 6.1700e-01 \\
    % 1.2500e+00 5.0100e-01 \\
    % 1.3750e+00  3.7500e-01 \\
    % 1.5000e+00 2.8800e-01 \\
    % 1.6250e+00 1.5500e-01 \\
    % 1.7500e+00  1.1900e-01 \\
    % 1.8750e+00  5.2659e-02 \\
    % 2.0000e+00 3.1566e-02 \\
    % 2.1250e+00  1.4603e-02 \\
    % 2.2500e+00  9.7238e-03 \\
    % 2.3750e+00  4.7526e-03 \\
    % 2.5000e+00  1.7638e-03 \\
    % };   
    % \addlegendentry{P=1, dscf metr. matlab} 
}
    \caption{\Gls{fer} of $(1024,512+32)$ under SCL, SCLF and PSCLF. and various designs of partitions $\mu$. }
    \label{fig:FER_mu}
\end{figure}

\autoref{fig:FER_mu} depicts the error-correction performance of \gls{psclf} with various designs of $\mu$.
For \gls{psclf}, two uniform structures $\mathcal{C}_{eq}=\{8,8,8,8\}$ for $P=4$ and $\mathcal{C}_{eq}=\{16,16\}$ for $P=2$ are used. 
The proposed approach based on the \glspl{cdf} is used to design two partitions $\mu$: $\mu=\{335,\textbf{409},589,\textbf{1023}\}$ \eqref{eq:mu_1dB}, and $\mu=\{410,\textbf{590},708,\textbf{1023}\}$ \eqref{eq:mu_2dB} that will be compared with the uniformly distributed partitions $\mu_{eq}=\{255,\textbf{511},767,\textbf{1023}\}$.
The boldface values compose the partition for $P=2$.

For $P=\{2,4\}$ partitions, the \gls{fer} is improved by using the proposed designs of partitions $\mu$ rather than $\mu_{eq}$. 
Overall, the \gls{fer} is better for $P=2$ than for $P=4$. 
Similar observations were made under \gls{pscf} \cite{PSCF} or \gls{pscl} \cite{partition_scl} decoding.
For $P=2$ and $P=4$, the error-correction performance of \gls{psclf} is better than that of \gls{sclf} when using $\mu$ \eqref{eq:mu_1dB}--\eqref{eq:mu_2dB} for $\text{FER}\geq3\cdot 10^{-5}$ and $\text{FER}\geq7\cdot 10^{-4}$, respectively.
If $\mu$ \eqref{eq:mu_1dB} is used with $P=2$ ($\mathbin{
\tikz[baseline]{\draw[color=matlab2,dashed,thick] (0pt,.5ex) -- (2ex,.5ex) -- (4ex,.5ex);
\draw[solid, color=matlab2,thick](2ex,.5ex) circle (.5ex);}}$), no loss is observed with respect to \gls{sclf}.
Up to $0.15$\,dB gain is observed at $\frac{E_b}{N_0}=2$\,dB if $P=2$ and $\mu$ designed for 2\,dB \eqref{eq:mu_2dB} ($\mathbin{
\tikz[baseline]{\draw[color=matlab2,thick] (0pt,.5ex) -- (2ex,.5ex) -- (4ex,.5ex);
\draw[solid, color=matlab2,thick](2ex,.5ex) circle (.5ex);}}$) which is compliant with our design approach.
For $P=4$, \gls{psclf} suffers from a $0.2$\,dB loss at $\frac{E_b}{N_0}=2.75$\,dB with respect to \gls{sclf}.
The crossover happens at $\frac{E_b}{N_0}=2.3$\,dB with $\mu$ \eqref{eq:mu_1dB}($\mathbin{
\tikz[baseline]{\draw[color=matlab1,dashed,thick] (0pt,.5ex) -- (2ex,.5ex) -- (4ex,.5ex);
\draw[solid, color=matlab1,thick] (1.5ex,0ex) -- (2.5ex,1ex);
\draw[solid, color=matlab1,thick] (1.5ex,1ex) -- (2.5ex,0ex);}}$) and $\mu$ \eqref{eq:mu_2dB}($\mathbin{
\tikz[baseline]{\draw[color=matlab1,solid,thick] (0pt,.5ex) -- (2ex,.5ex) -- (4ex,.5ex);
\draw[solid, color=matlab1,thick] (1.5ex,0ex) -- (2.5ex,1ex);
\draw[solid, color=matlab1,thick] (1.5ex,1ex) -- (2.5ex,0ex);}}$) and at $\frac{E_b}{N_0}=2.1$\,dB with $\mu_{eq}$ ($\mathbin{
\tikz[baseline]{\draw[color=matlab1,dotted,thick] (0pt,.5ex) -- (2ex,.5ex) -- (4ex,.5ex);
\draw[solid, color=matlab1,thick] (1.5ex,0ex) -- (2.5ex,1ex);
\draw[solid, color=matlab1,thick] (1.5ex,1ex) -- (2.5ex,0ex);}}$).
The error-correction performance is matched with \gls{scl}-16.

\subsection{Impact of the CRC Structure}\label{subsec:CRC_sim}
%Next, $\mu_{sim}$ \eqref{eq:mu_2dB} is used and $C=32$ is unchanged, however the distribution $\mathcal{C}$ of the \gls{crc} bits changes.
For $P=4$, simulations with \eqref{eq:mu_2dB} show a performance degradation starting at $\frac{E_b}{N_0}=2$\,dB.
As this behavior is not apparent for $P=2$  with $\mathcal{C}=\{16,16\}$, this is expected to be an impact of the \gls{crc} design.
Since the degradation occurs at higher $\frac{E_b}{N_0}$ values, the proposed \gls{crc} structure attributes more \gls{crc} bits to the last partition. %, i.e., $\mathcal{C}_2=\{7,7,7,11\}$.
According to \autoref{tab:prob_error}, at $\frac{E_b}{N_0}=2.75$\,dB, $56$\% of the first errors occurs in the last partition.
Among these errors, some will cause a collision with a \gls{crc} of 8 bits but not with a \gls{crc} of 11 bits.
Hence, by not passing the enlarged \gls{crc} code, flipping is conducted on this partition and the error may get corrected.

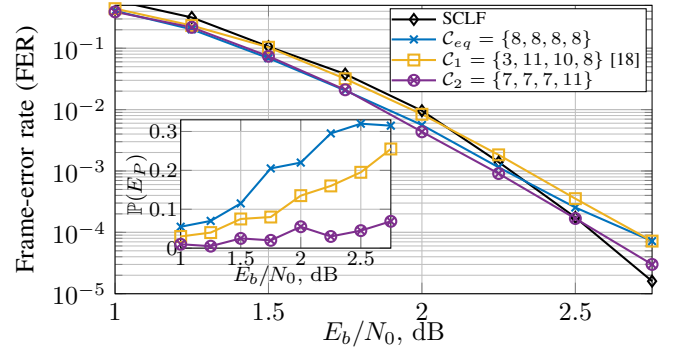
\begin{figure}[t!]
    \centering
    \vspace{2pt}
    \resizebox{.99\columnwidth}{!}{\begin{tikzpicture}
  \pgfplotsset{
    label style = {font=\fontsize{10pt}{8.2}\selectfont},
    tick label style = {font=\fontsize{10pt}{8.2}\selectfont}
  }

   \begin{semilogyaxis}[
    width=\columnwidth,
    height=0.62\columnwidth,
    xmin=1.0, xmax=2.75,
    xlabel={$E_b/N_0,\,\mathrm{dB}$},
    xlabel style={yshift=0.8em},
    ymin=1e-5, ymax=0.5,
    ylabel style={yshift=-0.1em},
    ylabel={Frame-error rate (FER)},
    yminorticks, xmajorgrids,
    ymajorgrids, yminorgrids,
    legend style={at={(0.99,0.99)},anchor=north east},
    %column font, row sep
    legend style={legend columns=1, font=\scriptsize, row sep=-1mm},
    legend style={fill=white, fill opacity=1, draw opacity=1,text opacity=1}, 
    legend cell align={left},
    legend style={inner xsep=0.2pt, inner ysep=-1pt}, % TIGHTER
    mark size=1.6pt, mark options=solid,
    ]   

       \addplot[color=black, mark=diamond, line width=0.8pt, mark size=2.1pt]
table[x=xdata,y=ydata]{data/data_FER/sclf.data}; 
    \addlegendentry{SCLF}  

\addplot[color=matlab1,mark=x, line width=0.8pt, mark size=2.1pt]
   table[x=xdata,y=ydata]{data/data_CRC/CRC_8888.data}; 
    \addlegendentry{$\mathcal{C}_{eq}=\{8,8,8,8\}$}

   \addplot[color=matlab3,mark=square, line width=0.8pt, mark size=2.1pt]
   table[x=xdata,y=ydata]{data/data_CRC/CRC_311108.data}; 
    \addlegendentry{$\mathcal{C}_1=\{3,11,10,8\}$ \cite{CRC_PSCL_virtual}}
    
       \addplot[color=matlab4,mark=otimes, line width=0.8pt, mark size=2.1pt]
   table[x=xdata,y=ydata]{data/data_CRC/CRC_77711.data}; 
    \addlegendentry{$\mathcal{C}_2=\{7,7,7,11\}$}
  \end{semilogyaxis}

  \begin{axis}[
    axis background/.style={fill=white, opacity=1},
    at={(0.1\columnwidth,(0.07\columnwidth)},
    width=0.5\columnwidth,
    height=0.375\columnwidth,
    opacity=1,
    xmin=1.0, xmax=2.75,
    xlabel={$E_b/N_0,\,\mathrm{dB}$},
    xlabel style={yshift=1.2em,font=\footnotesize},
    ymin=0, ymax=0.33,
    ytick={0,0.1,0.2,0.3},
    ylabel={$\mathbb{P}(E_P)$},
    ylabel style={yshift=-1.8em,font=\footnotesize},
    yminorticks, xmajorgrids,
    ymajorgrids, yminorgrids,
    yticklabel style = {font=\footnotesize,xshift=0.5ex},
    xticklabel style = {font=\footnotesize,yshift=0.5ex},
    legend style={at={(0.97,0.97)},anchor=north east},
    %column font, row sep
    legend style={legend columns=3, font=\scriptsize, row sep=-1mm},
    legend style={fill=white, fill opacity=1, draw opacity=1,text opacity=1}, % for future use maybe ? %opacity of filling/border and inside text
    legend style={inner xsep=0.2pt, inner ysep=-1pt}, % TIGHTER
    mark size=1.6pt, mark options=solid,
    ]
    \addplot[color=matlab1,mark=x, line width=0.8pt, mark size=2.1pt]
   table[x=xdata,y=ydata]{data/data_collision/collision_8888.data}; 

   \addplot[color=matlab3,mark=square, line width=0.8pt, mark size=2.1pt]
   table[x=xdata,y=ydata]{data/data_collision/collision_311108.data}; 
       \addplot[color=matlab4,mark=otimes, line width=0.8pt, mark size=2.1pt]
   table[x=xdata,y=ydata]{data/data_collision/collision_77711.data}; 
    \end{axis}
\end{tikzpicture}}
    \caption{\gls{fer} of $(1024,512+32)$ with multiple \gls{crc} structures. }
    \label{fig:FER_CRC}
\end{figure}

\autoref{fig:FER_CRC} shows an error-correction performance comparison under \gls{psclf} decoding for various \gls{crc} structures. 
The curve for \gls{sclf} decoding is provided as a reference.
It can be seen that at a \gls{fer} of $10^{-4}$, using $\mathcal{C}_2=\{7,7,7,11\}$ ($\mathbin{
\tikz[baseline]{\draw[color=matlab4,thick] (0pt,.5ex) -- (2ex,.5ex) -- (4ex,.5ex);
\draw[solid, color=matlab4,thick](2ex,.5ex) circle (.6ex);
\draw[solid, color=matlab4,thick] (1.5ex,0ex) -- (2.5ex,1ex);
\draw[solid, color=matlab4,thick] (1.5ex,1ex) -- (2.5ex,0ex);}}$) provides a gain of over $0.15$\,dB compared to $\mathcal{C}_{eq}$.
From the same figure, it can be seen that the \gls{crc} structure $\mathcal{C}_1=\{3,11,10,8\}$ for a $(1024,512+32)$ polar code from \cite{CRC_PSCL_virtual} ($\mathbin{
\tikz[baseline]{ \draw[color=matlab3,-,thick] (0pt,.5ex) -- (2ex,.5ex) -- (4ex,.5ex);
\draw[color=matlab3,-,thick] (1.5ex,.0ex) rectangle (2.5ex,1ex);}}$) suffers from error-correction degradation in both low and high \gls{snr}. 
The first partition is protected by only 3 \gls{crc} bits, we tracked that there was 3 times more collision occurring in the first partition at low \gls{snr} in comparison with the performance by using $\mathcal{C}_2=\{7,7,7,11\}$.
As depicted in \autoref{fig:FER_CRC}, the collision in the last partition also increases with $\mathbb{P}(E_P)=25.5$\% for  $\frac{E_b}{N_0}=2.75$\,dB. 
The reason is that the second and third partitions have a small error probability but are protected by a large number of \gls{crc} bits, respectively, $11$ and $10$.
This limits the number of \gls{crc} bits in the last partition, facing most of the errors.

The probability $\mathbb{P}(E_P)$ that the decoding error is a collision in the last partition is also depicted in \autoref{fig:FER_CRC}.
As described previously, by using $\mathcal{C}_{eq}$, the $\mathbb{P}(E_P)$ increases with $\frac{E_b}{N_0}$ and reaches a $\mathbb{P}(E_P)=31.5$\% for $\frac{E_b}{N_0}=2.75$\,dB.
Our proposed design reduces the collision $\mathbb{P}(E_P)$ to $6.8$\% for $\frac{E_b}{N_0}=2.75$\,dB.%, i.e., only $6.8$\% of the decoding errors were uncorrectable at that \gls{snr}.
\subsection{Partitioned SCLF with $N=512$ and $N=2048$}
\begin{figure}[t!]
    \centering
    \vspace{2pt}
    \resizebox{.99\columnwidth}{!}{\begin{tikzpicture}
  \pgfplotsset{
    label style = {font=\fontsize{10pt}{8.2}\selectfont},
    tick label style = {font=\fontsize{10pt}{8.2}\selectfont}
  }

   \begin{semilogyaxis}[
    width=\columnwidth,
    height=0.60\columnwidth,
    xmin=1.0, xmax=3.25,
    xlabel={$E_b/N_0,\,\mathrm{dB}$},
    xlabel style={yshift=0.8em},
    ymin=2e-5, ymax=0.5,
    ylabel style={yshift=-0.1em},
    ylabel={Frame-error rate (FER)},
    yminorticks, xmajorgrids,
    ymajorgrids, yminorgrids,
    legend style={at={(0.01,0.01)},anchor=south west},
    %column font, row sep
    legend style={legend columns=1, font=\scriptsize, row sep=-1mm},
    legend cell align={left},
    legend style={fill=white, fill opacity=1, draw opacity=1,text opacity=1}, % for future use maybe ? %opacity of filling/border and inside text
    legend style={inner xsep=0.2pt, inner ysep=-1pt}, % TIGHTER
    mark size=1.6pt, mark options=solid,
    ]   

       \addplot[color=black, dashed, mark=square, line width=0.8pt, mark size=2.1pt]
    table[x=Eb,y=FER]{data/2048_1024_09/SCL_16.data};
    \addlegendentry{SCL-16}    
    \addplot[color=black, dashed, line width=0.8pt, mark size=2.1pt]
    table[x=Eb,y=FER]{data/2048_1024_09/SCL_4.data};    
    \addlegendentry{SCL-4} 
       \addplot[color=black, mark=diamond, line width=0.8pt, mark size=2.1pt]
table[x=SNR,y=FER]{data/2048_1024_09/SCLF.data}; 
%table[x=SNR,y=FER]{data/data_FER/2k_32.data}; 
    \addlegendentry{SCLF}     

  \addplot[color=matlab1,dotted, mark=x, line width=0.8pt, mark size=2.1pt]
table[x=SNR,y=FER]{data/2048_1024_09/PSCLF_mu_eq_8888.data}; 
%table[x=SNR,y=FER]{data/data_FER/2k_mu_eq_8888.data}; 
    \addlegendentry{PSCLF, $\mu_{eq}$ - $\mathcal{C}_{eq}$}

  \addplot[color=matlab1, mark=x, line width=0.8pt, mark size=2.1pt]
table[x=SNR,y=FER]{data/2048_1024_09/PSCLF_mu_8888.data}; 
%table[x=SNR,y=FER]{data/data_FER/2k_mu_8888.data}; 
    \addlegendentry{PSCLF, $\mu$ - $\mathcal{C}_{eq}$}

      \addplot[color=matlab4, mark=otimes, line width=0.8pt, mark size=2.1pt]
table[x=SNR,y=FER]{data/2048_1024_09/PSCLF_mu_77711.data};
%table[x=SNR,y=FER]{data/data_FER/2k_mu_77711.data}; 
    \addlegendentry{PSCLF, $\mu$ - $\mathcal{C}_2$}

       \addplot[color=black, dashed, mark=square, line width=0.8pt, mark size=2.1pt]
    table[x=Eb,y=FER]{data/512_256_075/SCL_16.data};
    % ------------------------------------------
    \addplot[color=black, dashed,  line width=0.8pt, mark size=2.1pt]
    table[x=Eb,y=FER]{data/512_256_075/SCL_4.data};
    %-------------------------
       \addplot[color=black,mark=diamond, line width=0.8pt, mark size=2.1pt]
table[x=SNR,y=FER]{data/data_FER/512_32.data};     

  \addplot[color=matlab1,mark=x, line width=0.8pt, mark size=2.1pt]
table[x=SNR,y=FER]{data/data_FER/512_mu_8888.data}; 

      \addplot[color=matlab4, mark=otimes, line width=0.8pt, mark size=2.1pt]
table[x=SNR,y=FER]{data/data_FER/512_mu_77711.data}; 

      \addplot[color=matlab1, dotted, mark=x, line width=0.8pt, mark size=2.1pt]
table[x=SNR,y=FER]{data/data_FER/512_mu_eq_8888.data};

\draw[dashed]  (axis cs:2.35,0.02) ellipse [    x radius = 5, y radius = 1.2,rotate=5];
\node [rotate=-20]at (axis cs:2.3,0.125) {\scriptsize $N=512$};
\draw[dashed]  (axis cs:1.45,0.06) ellipse [    x radius = 5, y radius = 1.2,rotate=-45];
\node [rotate=-35]at (axis cs:1.35,0.015) {\scriptsize $N=2048$};
  \end{semilogyaxis}

\end{tikzpicture}%

% \addplot[color=black,dashed, mark = x,  line width=0.8pt, mark size=2.1pt]
 %    table[row sep=crcr]{%
 %    1.0000    7.4700e-01  \\
 %    1.1250     6.1600e-01 \\
 %    1.2500    5.1500e-01 \\
 %    1.3750     3.8300e-01 \\
 %    1.5000      3.0300e-01 \\
 %    1.6250     1.8900e-01 \\
 %    1.7500     1.4600e-01\\
 %    1.8750     7.7399e-02  \\
 %    2.0000    4.5086e-02  \\
 %    2.1250    3.5984e-02 \\
 %    2.2500     1.4162e-02\\
 %    2.3750   9.7314e-03 \\
 %    2.5000   4.6904e-03  \\
 %    2.6250   2.1277e-03  \\  
 %    2.75  1.1289e-03  \\
 %    2.875  4.0902e-04  \\
 %    3.0  1.7054e-04 \\
 %    };   
 %    \addlegendentry{P=1 orig. metr. matlab} 

    % \addplot[color=red,dashed,  mark=x, line width=0.8pt, mark size=2.1pt]
    % table[row sep=crcr]{%   
    % 1.0000e+00 7.4200e-01 \\
    % 1.1250e+00 6.1700e-01 \\
    % 1.2500e+00 5.0100e-01 \\
    % 1.3750e+00  3.7500e-01 \\
    % 1.5000e+00 2.8800e-01 \\
    % 1.6250e+00 1.5500e-01 \\
    % 1.7500e+00  1.1900e-01 \\
    % 1.8750e+00  5.2659e-02 \\
    % 2.0000e+00 3.1566e-02 \\
    % 2.1250e+00  1.4603e-02 \\
    % 2.2500e+00  9.7238e-03 \\
    % 2.3750e+00  4.7526e-03 \\
    % 2.5000e+00  1.7638e-03 \\
    % };   
    % \addlegendentry{P=1, dscf metr. matlab} 
}
    \caption{\gls{fer} of $(2048,1024+32)$ and $(512,256+32)$ with $\mu_{eq}$ and our proposed $\mu$. CRC structures $\mathcal{C}_{eq}$ and $\mathcal{C}_2$ are used.}
    \label{fig:FER_2K}
\end{figure}
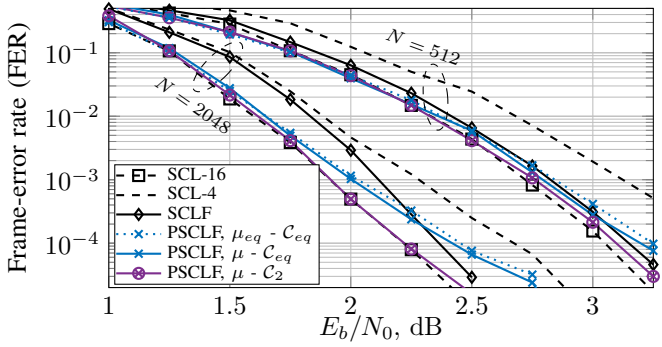

For $(2048,1024+32)$ (respectively $(512,256+32)$), Figure \ref{fig:FER_2K} depicts the error-correction performance of PSCLF with our proposed $\mu=\{819,1181,1417,2047\}$ $\left(\mu=\{199,279,344,511\}\right)$ and proposed CRC structure $\mathcal{C}_2=\{7,7,7,11\}$ with respect to $\mu_{eq}=\{511,1023,1535,2047\}$ $\left(\mu_{eq}=\{127,255,383,511\}\right)$ and $\mathcal{C}_{eq}=\{8,8,8,8\}$.
For both codes, using the designed partition $\mu$ improves the performance, while the pattern of CRC bits $\mathcal{C}_2=\{7,7,7,11\}$ improves even more the performance.
We observe up to $0.2$ dB at $\text{FER}=10^{-4}$ for $N=2048$.
For both codes, the decoding performance of PSCLF is matched with SCL-16.

\subsection{Average Execution Time of \gls{psclf}}
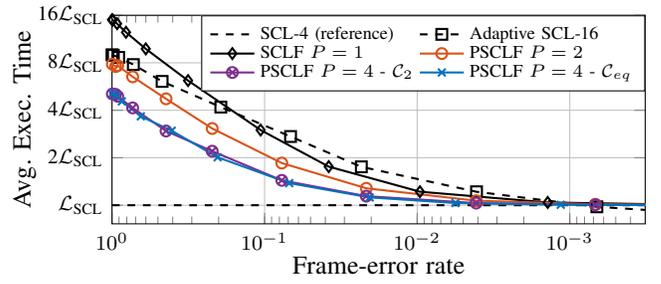
\begin{figure}[t!]
    \centering
    \resizebox{.99\columnwidth}{!}{\usetikzlibrary{spy}
\begin{tikzpicture}[spy using outlines={circle, magnification=2, connect spies}]
  \pgfplotsset{
    label style = {font=\fontsize{10pt}{8.2}\selectfont},
    tick label style = {font=\fontsize{8pt}{8.2}\selectfont}
  }

   \begin{semilogyaxis}[%
    width=\columnwidth,
    height=0.5\columnwidth,
    xmin=3.2e-04, xmax=1e-00,
    xlabel={Frame-error rate},
    xlabel style={yshift=0.4em},
    ymin=2000, ymax=41986,
    x dir=reverse,
    xmode=log,
    ytick = {2624,5248,10496,20992, 41986},
    yticklabels={$\latSCL$,2$\latSCL$,4$\latSCL$,8$\latSCL$,16$\latSCL$},
    ylabel style={yshift=-0.1em},
    ylabel={Avg. Exec. Time},
    xlabel style={yshift=0.2em},
    yminorticks, xmajorgrids,
    ymajorgrids, yminorgrids,
    legend style={at={(1.0,1.0)},anchor=north east},
    legend style={legend columns=2, font=\scriptsize, column sep=0mm, row sep=-1mm}, 
    legend cell align={left},
    mark size=1.8pt, mark options=solid,
    ]

  % \addplot [mark=diamond, color=black]table[row sep=crcr]{
  % 0   10240\\      
  % 3   10240\\
  % };  
  % \addlegendentry{SC ref.}
  
  % \addplot [mark=x, color=MyBlue]table[row sep=crcr]{
  % 0   153600\\
  % 0.25   153456.64\\
  % 0.5   152404.992\\
  % 0.75   149060.608\\
  % 1   135092.224\\
  % 1.25   114484.224\\
  % 1.5   83119.104\\
  % 1.75   54836.9408\\
  % 2   32311.3984\\
  % 2.25   18960.41472\\
  % 2.5   13318.656\\
  % 2.75   11196.010496\\
  % 3   10509.01504\\
  % };
  % \addlegendentry{SCF}

  \addplot [dashed, color=black, line width=0.8pt, mark size=2.1pt]table[row sep=crcr]{
  3.2e-04   2624\\      
  1e-00   2624\\
  };  
  \addlegendentry{SCL-4 (reference)}

  \addplot [mark=square,dashed, color=black, line width=0.8pt, mark size=2.1pt]table[x=FER,y=latency]{data/Adaptive_SCL16_1k.data};  
  \addlegendentry{Adaptive SCL-16}

\addplot [mark=diamond, color=black, line width=0.8pt, mark size=2.1pt]
  table[x=xdata,y=ydata]{data/data_avg_exe_time/P1_crc_32.data};
  \addlegendentry{SCLF $P=1$}

  \addplot[color=matlab2, solid,mark=o, line width=0.8pt, mark size=2.1pt]
  table[x=xdata,y=ydata]{data/data_avg_exe_time/P2_crc_1616.data};
  \addlegendentry{PSCLF $P=2$}   

    \addplot [color=matlab4,mark=otimes, line width=0.8pt, mark size=2.1pt]
  table[x=xdata,y=ydata]{data/data_avg_exe_time/P4_crc_77711_ET.data};
  \addlegendentry{PSCLF $P=4$ - $\mathcal{C}_2$}   

  \addplot [color=matlab1,mark=x, line width=0.8pt, mark size=2.1pt]
  table[x=xdata,y=ydata]{data/data_avg_exe_time/P4_crc_8888.data};
  \addlegendentry{PSCLF $P=4$ - $\mathcal{C}_{eq}$}   

   % \coordinate (spypoint) at (axis cs:0.0005,2600); %xy to spy
   % \coordinate (magnifyglass) at (axis cs:0.002,7000); %sy to zoom
  \end{semilogyaxis}

  %\spy [gray, size=1.5cm] on (spypoint) in node[fill=white] at (magnifyglass);

\end{tikzpicture}}
    \caption{Average execution time of \gls{sclf}, \gls{psclf}, and decoding latency of SCL $\latSCL$ for $(1024,512+32)$ code.}
    \label{fig:exec_psclf}
\end{figure}

\gls{psclf} improves the decoding performance while reducing the average execution time.
This reduction is analysed in this section.
%Complexity reduction was observed with another metric for \gls{pscl} in \cite{segmented_scl} and \gls{pscf} in \cite{PSCF}.
All average execution-time curves depicted in \autoref{fig:exec_psclf} are computed based on Section\,\ref{subsec:exec_psclf}, by choosing $\varphi=64$ processing elements in one \gls{sc} module as selected in \cite{PolarBear}.
The parallel list implementation is used, i.e., $L=4$ \gls{sc} modules are considered for PSCLF.
In \autoref{fig:FER_mu}, \gls{ascl}-16 is shown to have similar error-correction performance.
For \gls{ascl}-16, the use of $4$ parallel SC modules is simulated for consistency in terms of area.
The average execution time of \gls{ascl} and \gls{sclf} verify $\avgASCL\geq \latSC$ and $\avgSCLF\geq \latSCL$ with equality if no additional decoding trial is performed.
For FER greater than $10^{-3}$, the average execution times follow $\avgPSCLF<\avgASCL$.

\autoref{fig:exec_psclf} depicts $\avgSCLF$ and the reduction $\avgPSCLF$ by using our proposed partitioned approach.
%By tracking the average number of flips in each partition, the average execution time of \gls{psclf} is shown to be smaller than that of \gls{sclf}.
At $\text{FER}=10^{-1}$, the reduction is of $29$\% and $45$\% for $P=2$ and $P=4$, respectively.
At $\text{FER}=10^{-2}$, the reduction is now of $5$\% and $14$\% for $P=2$ and $P=4$, respectively.
The gain is especially visible in undecidable areas where the \gls{psclf} decoding algorithm is early terminating, i.e., does not reach the last partition.
By having more partitions, the size of each partition reduces which mitigates the impact of flipping, hence $\avgPSCLF$ with $P=4$ is smaller than with $P=2$.
%It leads to a trade-off between error-correction performance and average execution time.
The \gls{crc} structure has no impact on $\avgPSCLF$ such that the proposed \gls{crc} structure design should be prioritised over the uniformly distributed \gls{crc} structure $\mathcal{C}_{eq}$.

\subsection{Future Research Directions}
%PSCLF reduces the average execution time while improving the error-correction performance.
%However, the total number of \gls{crc} bits $C$ is large and a proposed \gls{crc} structure is required to avoid false positives.
Reducing the total number of \gls{crc} bits $C$ will improve performance for \gls{sclf} ($P=1$). 
It would be interesting to carry out error-correction and average execution-time comparisons between \gls{psclf} ($P\geq2$) and \gls{sclf} embedding complexity reduction techniques, e.g., a restart mechanism \cite{GRM_TSP}.
%However, it does not exist yet for \gls{sclf}.
\section{Conclusion}
In this paper, we proposed the \acrfull{psclf} decoding algorithm.
In \gls{psclf}, the code is broken into partitions and each partition is decoded with a \gls{crc}-aided SCLF decoder.
Numerical results show that the proposed \gls{psclf} algorithm gains up to 0.25\,dB with respect to \gls{sclf} in terms of \gls{fer}.
To obtain this gain while using a low-complexity flip metric, we proposed a design of partitions and a \gls{crc} structure both tailored to \gls{psclf}. 
%The partition design permits a gain of $0.2$ dB at $\text{FER}=10^{-3}$ with 4 partitions. 
%The partitions are distributed such that the probability of errors in each partition is even.
The \gls{crc} design mitigates the impact of \gls{crc} collisions in \gls{psclf}.
At a \gls{fer} of $10^{-4}$, the proposed \gls{crc} structure was shown to offer a gain of 0.2\,dB over the regular \gls{crc} structure.
The average execution time of \gls{psclf} was estimated to be 1.5 times lower than that of \gls{sclf}.
%A large gain is brought by the possibility for \gls{psclf} to early-terminate while both decoding algorithm have a similar average execution time for low \gls{fer}.
\bibliographystyle{IEEEtran}
\bibliography{IEEEabrv,ConfAbrv,references}

\begin{thebibliography}{10}
\providecommand{\url}[1]{#1}
\csname url@samestyle\endcsname
\providecommand{\newblock}{\relax}
\providecommand{\bibinfo}[2]{#2}
\providecommand{\BIBentrySTDinterwordspacing}{\spaceskip=0pt\relax}
\providecommand{\BIBentryALTinterwordstretchfactor}{4}
\providecommand{\BIBentryALTinterwordspacing}{\spaceskip=\fontdimen2\font plus
\BIBentryALTinterwordstretchfactor\fontdimen3\font minus
  \fontdimen4\font\relax}
\providecommand{\BIBforeignlanguage}[2]{{%
\expandafter\ifx\csname l@#1\endcsname\relax
\typeout{** WARNING: IEEEtran.bst: No hyphenation pattern has been}%
\typeout{** loaded for the language `#1'. Using the pattern for}%
\typeout{** the default language instead.}%
\else
\language=\csname l@#1\endcsname
\fi
#2}}
\providecommand{\BIBdecl}{\relax}
\BIBdecl

\bibitem{ArikanPolarCodes}
E.~Ar{\i}kan, ``Channel polarization: a method for constructing
  capacity-achieving codes for symmetric binary-input memoryless channels,''
  \emph{{IEEE} Trans. Inf. Theory}, vol.~55, no.~7, pp. 3051--3073, Jul. 2009.

\bibitem{SCL}
I.~Tal and A.~Vardy, ``List decoding of polar codes,'' \emph{{IEEE} Trans. Inf.
  Theory}, vol.~61, no.~5, pp. 2213--2226, Mar. 2015.

\bibitem{scf_intro}
O.~Afisiadis, A.~Balatsoukas-Stimming, and A.~Burg, ``A low-complexity improved
  successive cancellation decoder for polar codes,'' in \emph{Asilomar Conf. on
  Signals, Syst., and Comput. (ACSSC)}, Nov. 2014.

\bibitem{standard}
$3^{\text{rd}}$ Generation Partnership Project~({3GPP}), ``Multiplexing and
  channel coding,'' \emph{3GPP 38.212 V.15.3.0}, 2018.

\bibitem{dyn_scf}
L.~Chandesris \emph{et~al.}, ``{Dynamic-SCFlip} decoding of polar codes,''
  \emph{{IEEE} Trans. Commun.}, vol.~66, no.~6, pp. 2333--2345, Jun. 2018.

\bibitem{partition_scl}
S.~Hashemi, A.~Balatsoukas-Stimming, P.~Giard \emph{et~al.}, ``Partitioned
  successive-cancellation list decoding of polar codes,'' in \emph{{IEEE} Int.
  Conf. on Acoustics, Speech, and Signal Process. ({ICASSP})}, 2016, pp.
  957--960.

\bibitem{segmented_scl}
H.~Zhou, C.~Zhang, W.~Song \emph{et~al.}, ``Segmented {CRC}-aided {SC} list
  polar decoding,'' in \emph{{{IEEE} Veh. Technol. Conf.}}, 2016.

\bibitem{PSCF}
F.~Ercan \emph{et~al.}, ``Partitioned successive-cancellation flip decoding of
  polar codes,'' in \emph{{IEEE} Int. Conf. Commun. (ICC)}, 2018, pp. 1--6.

\bibitem{first_SCLF}
Y.~Yongrun, P.~Zhiwen, L.~Nan \emph{et~al.}, ``Successive cancellation list
  bit-flip decoder for polar codes,'' in \emph{Int. Conf. on Wireless Commun.
  and Signal Process. ({WCSP})}, 2018.

\bibitem{flip_criteria_real_time}
F.~Cheng \emph{et~al.}, ``Bit-flip algorithm for successive cancellation list
  decoder of polar codes,'' \emph{IEEE Access}, vol.~7, pp. 58\,346--58\,352,
  2019.

\bibitem{flip_criteria_fixed}
M.~Rowshan and E.~Viterbo, ``Improved list decoding of polar codes by
  shifted-pruning,'' in \emph{{IEEE} Inf. Theory Workshop (ITW)}, 2019, pp.
  1--5.

\bibitem{metric_with_alpha}
Y.~Pan, C.~Wang, and Y.~Ueng, ``Generalized {SCL-Flip} decoding of polar
  codes,'' in \emph{{IEEE} Global Telecommun. Conf. (GLOBECOM)}, 2020.

\bibitem{reduced_complexity_metric}
F.~Ivanov, V.~Morishnik, and E.~Krouk, ``Improved generalized successive
  cancellation list flip decoder of polar codes with fast decoding of special
  nodes,'' \emph{J. of Commun. Netw.}, vol.~23, no.~6, pp. 417--432, 2021.

\bibitem{dyn_sclf}
Y.~Shen \emph{et~al.}, ``Dynamic {SCL} decoder with path-flipping for {5G}
  polar codes,'' \emph{{IEEE} Wireless Commun. Lett.}, vol.~11, no.~2, pp.
  391--395, 2022.

\bibitem{AdaptiveSCL}
B.~Li \emph{et~al.}, ``An adaptive successive cancellation list decoder for
  polar codes with cyclic redundancy check,'' \emph{{IEEE} Commun. Lett.},
  vol.~16, no.~12, pp. 2044--2047, Nov. 2012.

\bibitem{SCL_LLR}
A.~Balatsoukas-Stimming, M.~Parizi, and A.~Burg, ``{LLR}-based successive
  cancellation list decoding of polar codes,'' \emph{{IEEE} Trans. Signal
  Process.}, vol.~63, no.~19, pp. 5165--5179, 2015.

\bibitem{ASCL}
B.~Li, H.~Shen, and D.~Tse, ``An adaptive successive cancellation list decoder
  for polar codes with cyclic redundancy check,'' \emph{{IEEE} Commun. Lett.},
  vol.~16, no.~12, pp. 2044--2047, 2012.

\bibitem{CRC_PSCL_virtual}
H.~Zhou \emph{et~al.}, ``Segmented successive cancellation list polar decoding
  with tailored {CRC},'' \emph{J. Signal Process. Syst.}, vol.~91, pp.
  923--935, 2018.

\bibitem{5G_DCA_WCNC}
C.~Pillet \emph{et~al.}, ``On list decoding of {5G-NR} polar codes,'' in
  \emph{{IEEE} Wireless Commun. and Netw. Conf. ({WCNC})}, South Korea, May
  2020.

\bibitem{GRM_TSP}
I.~Sagitov, C.~Pillet, A.~Balatsoukas-Stimming \emph{et~al.}, ``Generalized
  restart mechanism for successive-cancellation flip decoding of polar codes,''
  \emph{Under review}, vol.~99, no.~99, pp. 1--12, 9999.

\bibitem{semi_par_sc}
C.~Leroux, A.~Raymond, G.~Sarkis \emph{et~al.}, ``A semi-parallel
  successive-cancellation decoder for polar codes,'' \emph{{IEEE} Trans. Signal
  Process.}, vol.~61, no.~2, pp. 289--299, Jan. 2013.

\bibitem{PolarBear}
P.~Giard, A.~Balatsoukas-Stimming, T.~M{\"u}ller \emph{et~al.}, ``{PolarBear}:
  A 28-nm {FD-SOI ASIC} for decoding of polar codes,'' \emph{{IEEE} Trans.
  Emerg. Sel. Topics Circuits Syst.}, vol.~7, no.~4, pp. 616--629, 2017.

\end{thebibliography}
\end{document}